\documentclass[letter]{aa} 

\usepackage{graphicx}
\usepackage{txfonts}
\usepackage{color}

\begin{document}

   \title{Formation of an ultra-diffuse galaxy in the stellar filaments of NGC~3314A: caught in act?}

   \author{Enrichetta Iodice
          \inst{1}
          \and
          Antonio La Marca\inst{1}
          \and
          Michael Hilker\inst{2}
          \and
          Michele Cantiello\inst{3}
          \and
          Giuseppe D'Ago\inst{4}
          \and
          Marco Gullieuszik\inst{5}
          \and
          Marina Rejkuba\inst{2}
          \and
          Magda Arnaboldi\inst{2}
          \and
          Marilena Spavone\inst{1}
          \and
          Chiara Spiniello\inst{6}
          \and
          Duncan A. Forbes\inst{7}
          \and
          Laura Greggio\inst{5}
          \and
          Roberto Rampazzo\inst{5}
          \and
          Steffen Mieske\inst{8}
          \and
          Maurizio Paolillo\inst{9}
          \and
          Pietro Schipani\inst{1}
          }

   \institute{INAF $-$ Astronomical Observatory of Capodimonte, Salita Moiariello 16, I-80131, Naples, Italy\\
              \email{enrichetta.iodice@inaf.it}
         \and
             European Southern Observatory, Karl-Schwarzschild-Strasse 2, D-85748 Garching bei Muenchen, Germany\\
             \and
             INAF-Astronomical Observatory of Abruzzo, Via Maggini, 64100, Teramo, Italy\\
             \and
             Instituto de Astrofísica, Facultad de Física, Pontificia Universidad Católica de Chile, Av. Vicu\~{n}a Mackenna 4860, 7820436 Macul, Santiago, Chile\\
             \and
             INAF $-$ Osservatorio Astronomico di Padova, Vicolo dell'Osservatorio 5, I-35122 Padova, Italy\\
             \and
             Department of Physics, University of Oxford, Denys Wilkinson Building, Keble Road, Oxford OX1 3RH, UK\\
             \and
             Centre for Astrophysics and Supercomputing, Swinburne University of Technology, Hawthorn, Victoria 3122, Australia\\
             \and
             European Southern Observatory, Alonso de Cordova 3107, Vitacura, Santiago, Chile\\
             \and
             University of Naples ``Federico II'', C.U. Monte Sant'Angelo, Via Cinthia, 80126, Naples, Italy\\
             }

   \date{Received ....; accepted ...}

 
  \abstract
{The VEGAS imaging survey of the Hydra~I cluster reveals an extended network of stellar filaments to the south-west of the spiral galaxy NGC~3314A. 
Within these filaments, at a projected distance of $\sim40$ kpc from the galaxy, we discover an ultra-diffuse 
galaxy (UDG) with a central surface brightness of $\mu_{0,g}\sim26$~mag arcsec$^{-2}$ and effective radius $R_e\sim3.8$~kpc. 
{   This UDG, named UDG~32, is one of the faintest and most diffuse low-surface brightness galaxies in the Hydra~I cluster. }
 Based on the available data, we cannot exclude that this object is just seen in projection on top of the stellar filaments, 
thus being instead a foreground or background UDG in the cluster. However, the clear spatial coincidence of  UDG~32 
with the stellar filaments of NGC~3314A suggests that it might have formed from 
the material in the filaments, becoming a detached, gravitationally bound system. 
In this scenario, the origin of UDG~32 depends on the nature of the stellar filaments in NGC~3314A, 
which is still unknown.
They could result from ram-pressure stripping or have a tidal origin.
In this letter, we focus on the comparison of the observed properties of the stellar filaments and UDG~32, 
and speculate about their possible origin. 
{  The relatively red colour ($g-r=0.54 \pm 0.14$~mag) of the UDG, similar to that of the disk in NGC~3314A, combined with an 
age older than 1~Gyr, and the possible presence of a few compact stellar systems, all point 
towards a tidal formation scenario inferred for the UDG~32. }

}

   \keywords{Galaxies: clusters: individual: Hydra~I - Galaxies: photometry - Galaxies: dwarf - Galaxies: formation}

   \maketitle
%

\section{Introduction}\label{sec:intro}


 UDGs have a special role in the realm of the low-surface brightness (LSB) universe.
They are empirically defined to be faint ($\mu_{0,g} \geq 24$ mag arcsec$^{-2}$) and diffuse ($R_e \geq 1.5$~kpc) objects,
with stellar masses similar to dwarf galaxies \citep[$\sim 10^7 - 10^8$ M$_{\odot}$,][]{vanDokkum2015}.
The dark matter (DM) content of  UDGs is one of the open and most debated issues. 
{ Recent literature has suggested that UDGs are dwarfs in terms of their DM halo \citep[][]{Lim2018,Prole2019b}.
However, a larger DM amount was inferred for several UDGs \citep[e.g.,][]{vanDokkum2019,Forbes2020a,Gannon2021}.
In contrast, some others UDGs with very low DM content have also been discovered \citep[e.g.,][]{vanDokkum2018,Collins2021}.}
This opened new questions on the whole framework of galaxy formation and whether long-lived 
dwarf galaxies should have a massive DM halo, unless they have { a tidal origin \citep{Lelli2015}.}

To date, a significant population of UDGs 
has been found in dense environments, such as clusters and groups of galaxies, as well as in the field \citep{Koda2015,Roman2017a,vanderBurg2017,Shi2017,Muller2018,Venhola2017,Prole2019a,Forbes2019,Janssens2019,Forbes2020b,Habas2020}.
Based on their color distribution, morphology, stellar population and globular cluster (GC) systems, 
a wide range of observed properties are 
found for UDGs, which do not fit in a single formation scenario \citep[][]{Leisman2017,Roman2017a,Ferre-Mateu2018,Martin-Navarro2019,Lim2020,Forbes2020a,Rong2020}.

Several formation channels have been proposed for UDGs. 
{ As the first detection of UDGs suggested a high DM content in a few of them, they have been termed “failed” galaxies, 
because they have lost gas supply at an early epoch, 
which prevented formation of normal, higher surface brightness systems, while they have large effective radii 
comparable to normal (MW-like) galaxies \citep[][]{vanDokkum2015}. 
%
UDGs with stellar masses and DM content consistent with dwarf galaxies could form as a consequence of anomalously 
high spins of DM halos \citep{Amorisco2016,Rong2017,Tremmel2019}, or kinematical heating of their stars 
induced by internal processes \citep[i.e. gas outflows associated with feedback,][]{diCintio2017}.} 
According to \citet[][]{Sales2020}, a population of “genuine” LSB galaxies, with UDG properties, forms in the field and 
later enters the cluster environment, while the so-called tidal-UDGs (T-UDG), stem from luminous galaxies and evolve 
into UDGs due to cluster tidal forces that remove their DM content. 
Tidally released material during galaxy interactions was also invoked to explain the formation of the DM-free tidal dwarf galaxies \citep[][]{Lelli2015,Duc2014,Ploeckinger2018}.
Finally, \citet[][]{Poggianti2019} suggested that the DM-free UDGs might form from ram pressure stripped (RPS) gas clumps
in the extended tails of infalling cluster galaxies. 

To date, on the observational side, there are few cases of UDGs clearly associated with tidal debris. 
In the sample of tidal dwarf galaxies (TDGs) studied by \citet[][]{Duc2014}, 
two objects have $R_e$ and $\mu_0$ consistent with being UDGs, are clearly associated with the tidal tails 
of the massive early-type galaxy NGC~5557. 
In this work the authors did not describe them as UDGs, because the specific name for this class of galaxies 
was introduced later \citep[][]{vanDokkum2015}.

The first evidence of UDGs formed through galaxy interactions was presented by \citet[][]{Bennet2018}.
They reported the discovery of two new UDGs { (NGC~2708-Dw1 and NGC~5631-Dw1)} 
that are probably associated with the stellar streams around them 
caused by the encounters with nearby massive galaxies.
\citet[][]{Muller2019} have investigated the possible connection between the two UDGs (NGC1052-DF2 and NGC1052-DF4) 
in the NGC1052 group with the LSB features (loops and stellar streams) detected in the intra-group space. 
They pointed out that no clear association can be made between NGC1052-DF2's origin and the tidal interaction in the group where it resides.
{ Recently, \citet[][]{Montes2020} have proposed that the existence of faint stellar tails found in the outskirts of NGC1052-DF4 
result from tidal stripping. This mechanism could have also removed a significant percentage of the DM, 
thus explaining the low DM content of this UDG.}


{  In this letter we report the discovery of a UDG in the Hydra~I cluster   
\citep[$51\pm6$~Mpc,][]{Christlein2003}} located within the newly detected faint stellar 
filaments of NGC~3314A. 
We have investigated whether this UDG and the filaments could be associated, therefore pointing towards a 
new observational evidence for UDG formation in galaxy interactions.

\section{A UDG candidate in the stellar filaments of NGC~3314A: UDG~32} \label{sec:3314}

\begin{figure*}
    \centering
    \includegraphics[width=18cm]{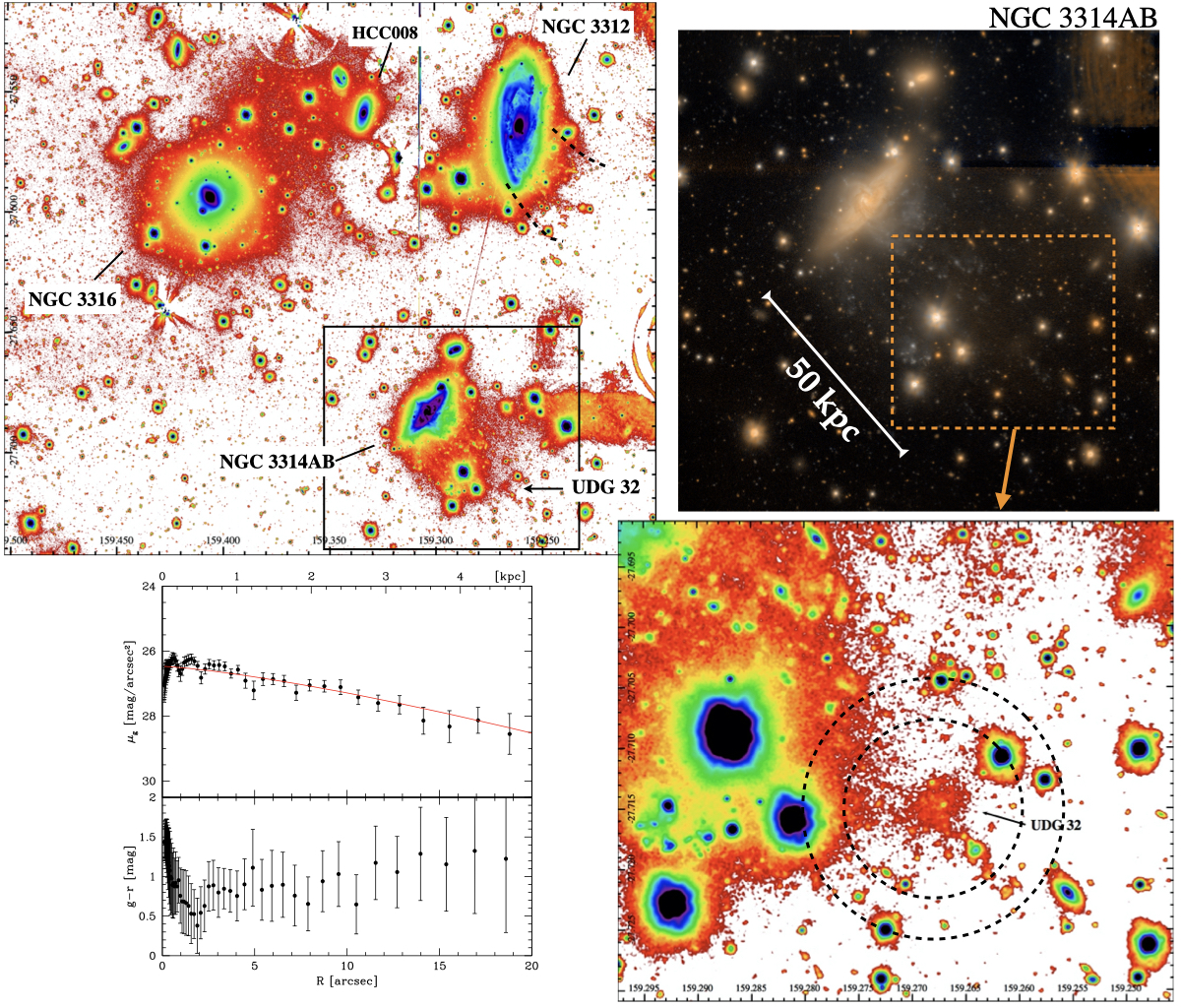}
    \caption{The top-left panel shows the extracted region of the VST mosaic ($16.2 \times 13.8$ arcmin), in the $g$ band,
    in the SE part of Hydra I cluster.
    The brightest galaxy members, NGC~3312, NGC~3314AB, NGC~3316 and HCC~008 are marked. 
    {  The dashed lines SW of NGC~3312 indicates the region where stellar filaments are detected.}
    The light of bright stars in the field has been modelled and subtracted from the reduced images {  \citep[see also][]{Iodice2020c}.}
    North is up and East is on the left.
    The black square is the region ($6.44 \times 6.04$~arcmin) around the galaxies NGC~3314AB shown as gr color composite 
    in the top-right panel. 
    {  The orange dashed box indicates the region around UDG~32 candidate, in the SW regions of NGC~3314AB, 
    described in the text and shown as $g$-band image ($4.8 \times 4.1$~arcmin) in the lower-right panel. 
    The dashed circular annulus indicates the region where we estimated the local background  (see text for details).
    In the lower right panel we show the azimuthally averaged surface brightness profile in the $g$ band of UDG~32, compared with the best fit Sersic law (red solid line). The  colour $g-r$ profile is shown in the lower-left sub-panels. }    }
    \label{fig:mosaic}
\end{figure*}

{   NGC~3314AB is a system of two spiral galaxies, member of the Hydra~I cluster (see top-left panel of 
Fig.~\ref{fig:mosaic}), seen in projection on top of each other along the line-of-sight. 
NGC~3314A is the foreground galaxy, with a heliocentric velocity of $cz=2795$ km~s$^{-1}$ \citep{Christlein2003}, 
while NGC~3314B has $cz=4665$ km~s$^{-1}$ \citep[][]{Keel2001,McMahon1992}.
Deep images of the Hydra\,I cluster, in the $g$ and $r$ bands, were acquired with the European Southern Observatory 
(ESO) VLT Survey Telescope (VST), as part of the {\it VST Early-type Galaxy Survey (VEGAS)}\footnote{see \url{http://www.na.astro.it/vegas/VEGAS/Welcome.html}} and presented in a recent paper by \citet{Iodice2020c}. 
The VST images have revealed an extended {  ($\sim 3.4$~arcmin $\sim 50$~kpc)}
network of stellar filaments in the SW direction of NGC~3314AB (see top-right panel of Fig.~\ref{fig:mosaic}).}
Since we do not see any discontinuity in the light distribution, {it is reasonable to associate} this structure with NGC~3314A. 

The WFPC2/HST images published by \citet[][]{Keel2001}, cover $63\arcsec \times 70\arcsec$, i.e. only $\sim17\%$ 
of the VST field around NGC~3314AB, and thus they do not show the whole extension of this system. 
In fact, the authors refer to the presence of "tails" in the SW regions of the disk in NGC~3314A.
%
{  We downloaded and analysed archival data for the ACS/HST mosaic, in the optical F606W filter, 
and GALEX images in FUV and NUV. They are briefly described in the Appendix~\ref{sec:arc_data}.
The highest signal-to-noise stacked ACS/HST mosaic (see Fig.\ref{fig:UV_HST}) 
extends out to $\sim 1.4$~arcmin ($\sim$ 20 kpc)
on the SW side. Here, the disk of NGC~3314A has a lop-sided morphology, and 
several filamentary structures, containing bright knots, are also detected. 
Since the ACS/HST mosaic covers the $\sim60\%$ of the VST images around NGC~3314, the morphological 
assessment of the outermost regions could not be done. 
Based on the new VST images, we concluded that the features detected from HST data on the SW side of NGC~3314A
are part of the more complex and extended structure of stellar filaments, which we estimated to be
 $\sim 20$ kpc wide and $\sim 50$ kpc long  (see top-right panel of Fig.~\ref{fig:mosaic}), 
becoming fainter ($\mu_g \sim 26-27$~mag/arcsec$^2$) at larger radii. }
%
%
The GALEX data show that they have  
UV emission out to $\sim3.7$~arcmin ($\sim54$ kpc) from the centre of NGC~3314A (see Fig.~\ref{fig:UV_HST}).

{  The HI surface brightness distribution presented by \citet[][]{McMahon1992} shows a long tail ($\sim30$~kpc) 
extending South of the galaxy centre and a clump to the SE. Both features spatially overlap with
the stellar filaments and the prominent star-forming regions in SE, detected 
in the VST images.}

 UDG~32\footnote{R.A.=159.26775 [deg], DEC=-27.715428 [deg]} 
is found within the stellar filaments of NGC~3314A, at a projected distance of 161.5~arcsec ($\sim 40$~kpc) 
from the  centre of NGC~3314A (see Fig.~\ref{fig:mosaic}). {  This object is outside 
the HST/ACS field of view. The GALEX images do not show any clear emission 
associated to UDG~32  (see Fig.~\ref{fig:UV_HST}). This might suggest that it does not have a 
UV emission, or, alternatively, the GALEX images are too shallow.
Multi-band observations for the Hydra I cluster are also available in the CFHT science archive, which are anyway
$\sim 2.5$~mag shallower in the $g$ and $r$ bands than VST data.
To date there are no further data as deep as VST images available, nor other studies of the region where we detected extended stellar filaments.}
%

\subsection{Detection and structural properties of UDG~32}\label{sec:color}

{  UDG~32 is one of the 20 newly discovered UDG candidates in the Hydra~I cluster.
The first sample of 12 UDGs detected by visual inspection were presented by \citet[][]{Iodice2020c}.
In order to increase the detection rate of the LSB galaxies in this cluster we have combined
the automatic tool {\sc SExtractor} \citep[][]{Bertin1996} with a subsequent visual inspection of 
the VST mosaics in the $g$ and $r$ bands. This led to the detection of 20 new UDGs.
While we plan to present a detailed study of UDGs in a forthcoming paper (La Marca et al. in preparation), which also
includes details on the detection process and UDGs selection, their properties are used here for comparison 
with one interesting UDG candidate, the UDG~32, which is the main focus of this Letter.

UDG~32 is one of the most diffuse UDGs detected in the Hydra~I cluster, with central 
surface brightness $\mu_0=26 \pm 1$~mag arcsec$^{-2}$ and $R_e=3.8\pm 1$~kpc. 
This object was found by visual inspection of the area around NGC~3314AB. Being so faint and diffuse, 
that it falls below the Sextractor detection threshold with parameters tuned for detection of LSBs in Hydra I 
( La Marca et al. in prep.).
Fig.~\ref{fig:corrRe} (right panel) shows that in the new sample of UDGs in Hydra~I there are few other 
UDGs with $\mu_0$ and $R_e$ similar to those of UDG~32, within the error estimate. 
Compared to the previous sample of LSB galaxies in Hydra~I by \citet[][]{Misgeld2008}, the faint tidally disrupting dwarf HCC~087 \citep[][]{Koch2012} is the only object falling in the parameter space occupied by UDGs.
In Appendix~\ref{sec:detection} we describe the main steps of the surface photometry performed in the area of NGC~3314AB, 
where UDG~32 is found, and how the structural parameters are derived.}  

{  UDG~32 has an integrated  colour of $g-r=0.54 \pm0.14$~mag and an absolute r-band magnitude of M$_r =-14.65$~mag.}
According to the colour-magnitude relation for early-type giant and dwarf galaxies in 
Hydra\,I \citep{Misgeld2008}, UDG~32 is consistent with being cluster members, and its
structural parameters are fully consistent with those measured for all UDGs in the cluster (see  Fig.~\ref{fig:corrRe}).
The azimuthally averaged surface brightness and colour profiles derived for UDG~32 are shown in Fig.~\ref{fig:mosaic}.
The redder  colours in the centre and the minimum in the surface brightness profile could be due to dust absorption. 
 Adopting the relation given by \citet{Into2013}\footnote{  This empirical relation was derived for the age range $10^6-10^{10}$~yrs and for metallicities between Z=0.0001 and Z=0.019 Z$_\odot$.}
the $g-r$ colour and M$_r$ are used to derive the stellar mass for UDG~32, which is $M_\ast \sim 9 \times 10^{7}$~M$_{\odot}$.

\begin{figure*}
   \includegraphics[width=9cm]{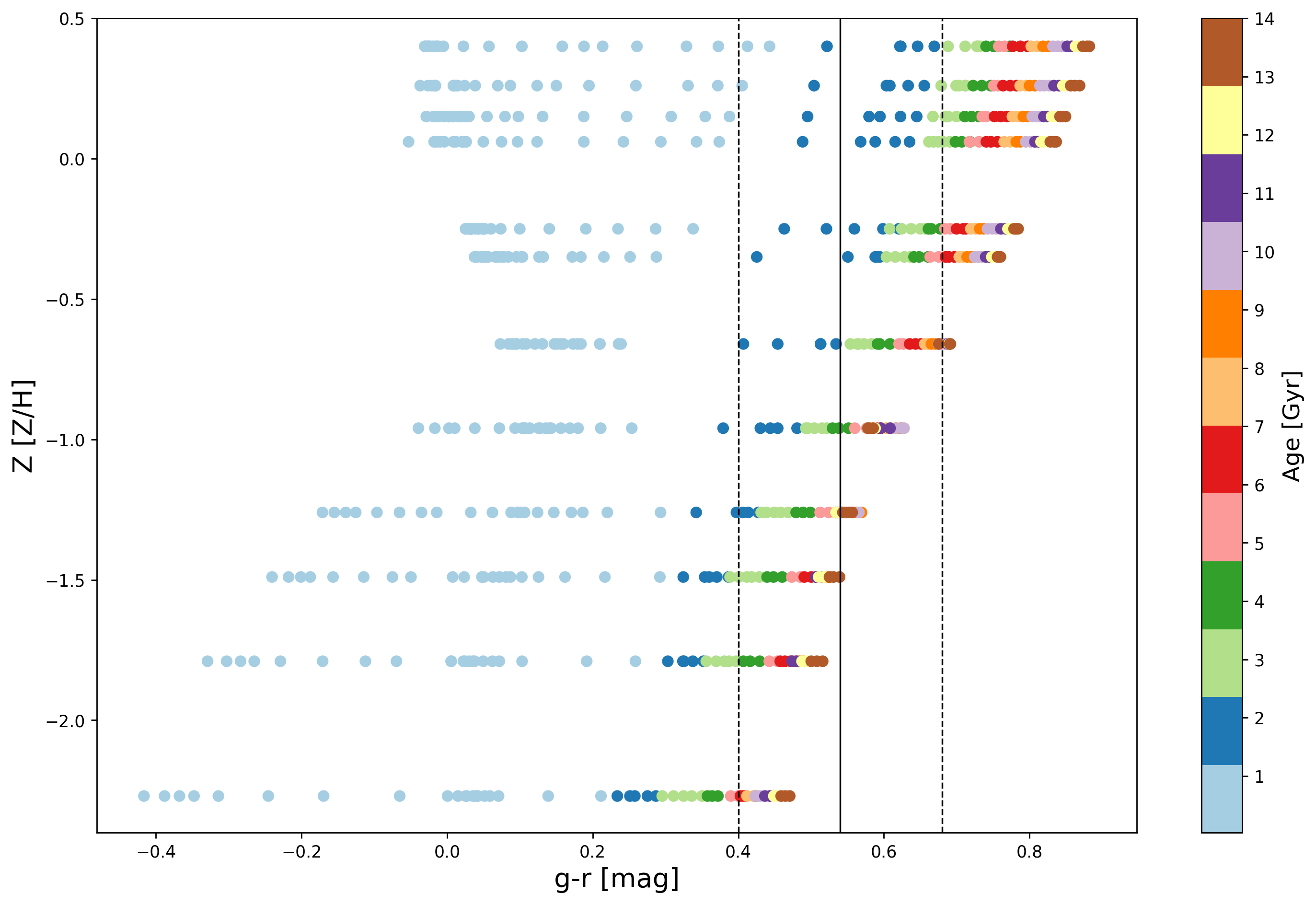} 
   \includegraphics[width=9cm]{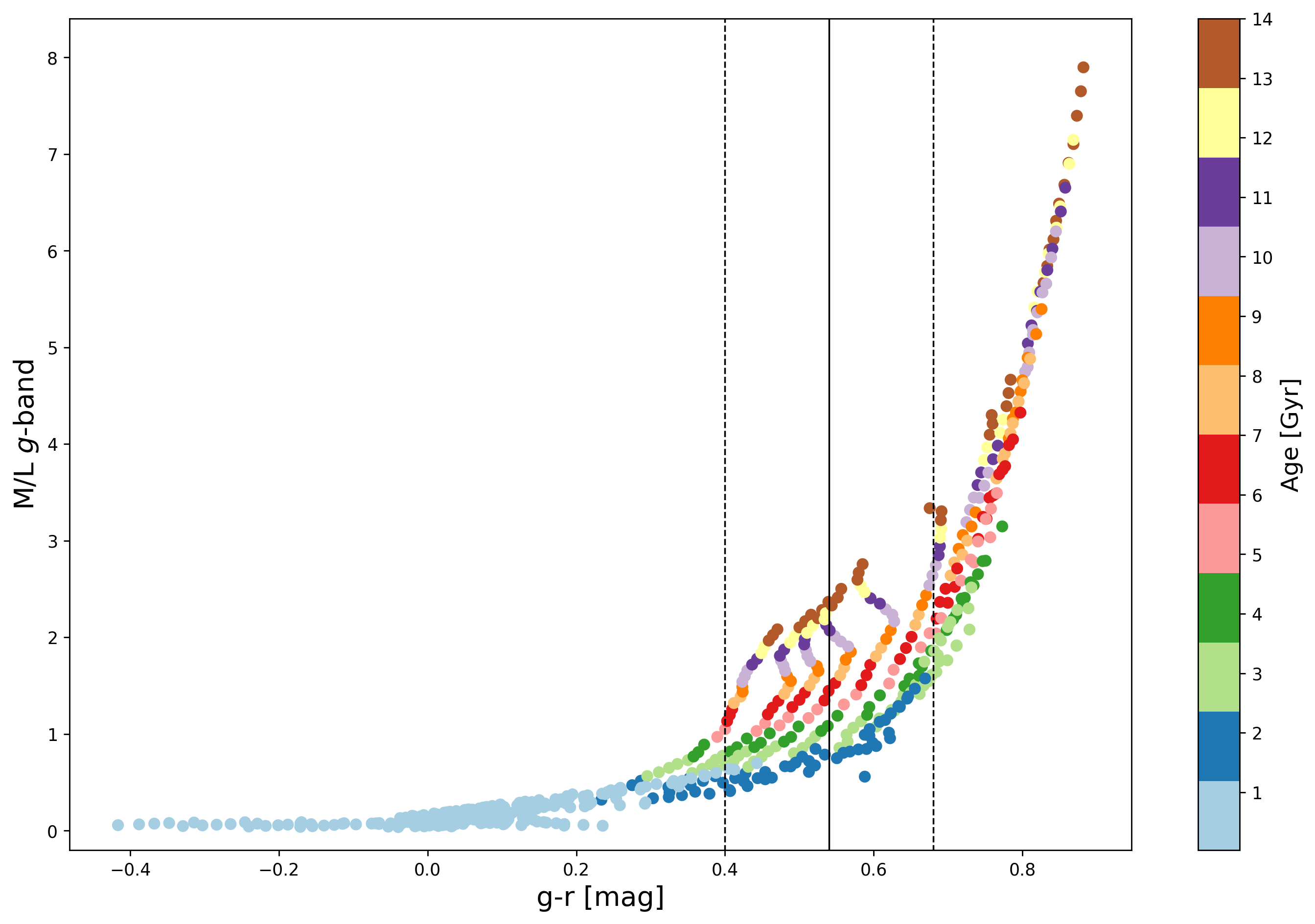}
    \caption{  E-MILES stellar population synthesis (SSP) models. 
    $g-r$ colors versus metallicity (left panel) and M/L ratio (right panel) predicted by the SSP models (colored points) corresponding to different ages, shown in colorbar on the right side. The solid black line indicates the integrated color measured for the UDG~32, and the two dashed lines corresponds to $1\sigma$ error. }
    \label{fig:age}
\end{figure*}

{  VST images in the region of NGC~3314A, are analysed to detect GC-like systems around UDG~32. To this aim, we have adopted  
the same procedure and assumptions to identify old GCs as in the first sample of UDGs in Hydra~I cluster 
\citep[see][]{Iodice2020c} and described in Appendix~\ref{sec:GC}.
We find the total number of GC candidates ($N_{GC}$) within $1.5R_e$, $3R_e$ and $5R_e$, to be $0\leq N_{GC}\leq 6$.  
In particular, we identified a compact source near the centre of the 
UDG~32 (at $\sim3.5\arcsec$), with colour, magnitude, and compactness properties close to those expected
for a bright old GC ($g-r\sim0.6$ mag, $m_g\sim24.9$ mag).

Based on the mean $g-r$ colour derived for UDG~32, within $1\sigma$ error, we used the 
E-MILES stellar population synthesis models \citep[][]{Vazdekis2012} to provide some constraints on the age and metallicity
for this object. Results are shown in Fig.~\ref{fig:age}, where the range of possible ages is reported considering 
all the possible metallicity values allowed by the stellar population models ($[Z/H]\sim-2.32$ and $0.22$ dex).
{  This analysis shows that populations with colors in the range $0.4 \leq g-r\leq0.64$ mag span a range of stellar ages
from 1 to 5 Gyrs for $-0.7 \leq [Z/H] \leq 0.5$. 
At lower metallicities ($-1.7\leq [Z/H] \leq -1.2$), older stellar ages are possible, up to 13 Gyrs.}
Assuming that UDG~32 is a normal LSB galaxy, according to the mass-metallicity relation 
derived for galaxies in the LSB regime \citep[Fig.~5 in][]{Pandya2018}, its stellar mass ($\sim 10^{8}$~M$_{\odot}$) 
would be consistent with a metallicity of $Z/H\geq -1.5$.
For this value, models shown in Fig.~\ref{fig:age} predict ages older than
3~Gyrs and a stellar mass-to-light ratio $M/L \geq 1.2$, in the $g$ band. 
%

However, it is worth noting that we cannot exclude that there is dust absorption in the UDG. 
Therefore, UDG~32 might be intrinsically bluer and, as consequence, even younger ages cannot be excluded 
without additional data.}

\begin{figure*}
\begin{tabular}{cc}
   \includegraphics[width=9cm]{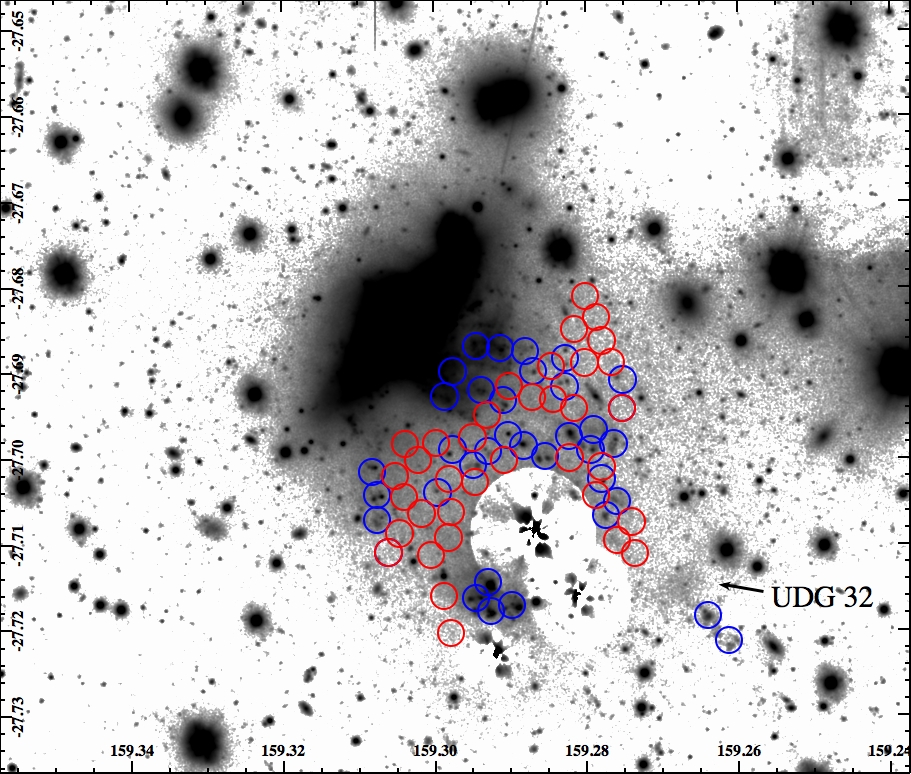} &
   \includegraphics[width=9cm]{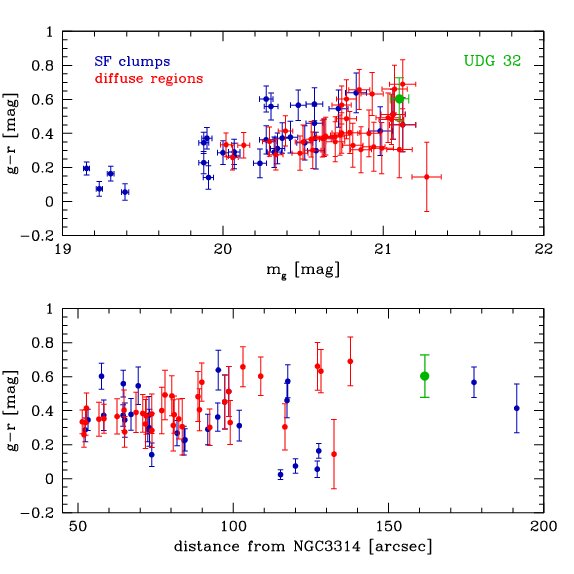}
   \end{tabular}
    \caption{Integrated  colours in the stellar filaments of NGC~3314A. 
    {\it Left panel -} Enlarged portion ($4.8 \times 4.1$~arcmin) of the Hydra I mosaic in the $g$ band centred on NGC~3314A, where the apertures adopted to compute the integrated  colours, shown in the right panel, are marked. {  The three bright stars located SW of the stellar filaments were modelled and subtracted from the image.}
    Blue and red open circles are for bright knots and diffuse regions, respectively. We used circular areas of 5.25 arcsec in radius. This was set on the most extended bright knots to enclose most of the emission. 
    {\it Right panel -}  Color-magnitude relation (upper panel) for the apertures covering the filaments in NGC~3314A 
    (shown in the left panel) and for the Hydra~I UDG~32 (green point). 
    In the lower panel, the integrated $g-r$  colours of the filament regions and of Hydra~I UDG32 are plotted as function of the projected distance from the galaxy  centre of NGC~3314A.
    {  Colors and magnitudes are not corrected for internal dust absorption.}}
    \label{fig:CMD_tails}
\end{figure*}

According to the hypothesis that UDG~32 is formed from the stellar material in the filaments of NGC~3314A, 
we would expect it to have a similar  colour as the galaxy. 
Therefore, we have derived the integrated $g-r$ colour in several regions along the filaments,
covering both diffuse emission regions and bright knots {  (see Fig.~\ref{fig:CMD_tails} and Appendix~\ref{sec:detection}).} 
 UDG~32  colour is fully consistent with the range of  colours in the filaments, which
is $0.3 \leq g-r \leq 0.65$~mag (see the top-right panel of Fig.~\ref{fig:CMD_tails}). 
On average, the bright knots in the filaments have bluer colours, with $0 \leq g-r \leq 0.5$~mag, therefore
these could be star-forming clumps. 
{  However, taking into account that the underlying diffuse emission and dust absorption
might affect the fluxes in the bright knots, bluer intrinsic colors in these regions can be also expected.} 
On average, the distribution of  colours in the filaments does not depend on the distance from the centre of NGC~3314A 
(see the lower-right panel of Fig.~\ref{fig:CMD_tails}).

%



\section{Discussion: formation scenarios} \label{sec:disc}
%
%
{  In this section, we discuss how the structural properties of UDG~32 fit to the proposed hypothesis on the 
association of this system with the stellar filaments of NGC~3314A. The spatial coincidence suggests that it might 
have formed from the  {baryons (gas and/or stars)} in the filaments, or it is a (foreground or background) UDG 
in the cluster, projected on top of the stellar filaments. 
The available data do not allow to resolve between these two equally viable formation channels.
%
However, due to the non-uniform distribution of the UDGs inside the Hydra~I cluster 
(see Fig.~\ref{fig:2Ddist}), 
the chance to find a UDG projected on top of a cluster member, is lower ($\sim 9$\%) in the area around NGC~3314A 
than in the cluster core ($\sim 50$\%). See Appendix~\ref{2D_density} for details. 

Under the hypothesis that  UDG~32 originated from the filaments, the formation mechanism depends on the nature of these structures.}
%

\subsection{On the nature of the stellar filaments in NGC~3314A: ram-pressure stripping versus tidal interaction}\label{sec:filaments_origin}
The morphology of the filaments and the presence of several bright, possibly star-forming knots resemble those seen in 
jelly-fish galaxies \citep[see][and references therein]{Poggianti2017} or, also, in spiral galaxies with 
unwinding arms \citep[][]{Bellhouse2021}.
{  In both cases, the filaments  {originate} from the RPS of the gas in the disk that subsequently formed new stars.}
Both the HI and radio continuum maps \citep[see figures 2 and 3 in][]{McMahon1992}  
show emissions elongated in the direction of stellar filaments in NGC~3314A.
RPS due to the interaction with the Hydra~I intra-cluster medium, that dominates the cluster core out to about 165~kpc in projected radius \citep[][]{Hayakawa2006}, would explain the alignment, under the assumption that NGC~3314A
is part of an infalling group into the cluster potential. 

In an alternative scenario, the disturbed morphology of NGC~3314A may be the result of a past tidal interaction 
with a nearby cluster member, as proposed by \citet[][]{McMahon1992}. To explain the asymmetries found in the HI distribution, 
the authors suggested that NGC~3314A and the giant spiral galaxy NGC~3312
could be members of a weakly interacting group in the foreground of the Hydra~I cluster.
NGC~3312 is a cluster member at a similar radial velocity as NGC~3314A ($\Delta V_{hel}=35$~km~s$^{-1}$), 
located SE of the cluster
core (see Fig.~\ref{fig:mosaic}). VST images show that the disk morphology of NGC~3312 is also quite disturbed: 
protruding stellar filaments are detected SW of the disk {  (see Fig.~\ref{fig:mosaic})}. 
Ongoing tidal interaction with the background galaxy of the system, NGC~3314B, could be reasonably excluded
since this galaxy has an undisturbed morphology and very different radial velocity \citep[see also][]{Keel2001}.
{  However, a high-speed encounter in the past, between the two galaxies, or with other cluster members, 
which induced a tidal distortion in the disk of NGC~3314A, cannot be excluded.
There are two more bright galaxies in the field, NGC~3316 and HCC~008, where the deep VST images reveal distorted morphology 
in the outskirts (see Fig.~\ref{fig:mosaic}). These are cluster members with larger radial velocities 
\citep[$cz\sim4000$~km/s][]{Christlein2003}. 
The intra-cluster region between NGC~3312 and NGC~3314A is strongly affected by the residual light from the two bright stars 
in the cluster core, modelled and removed from the image \citep[see][]{Iodice2020c}, which limits the
detection of other LSB features, as possible remnants of past interaction between the two galaxies. }

%
%
%
%
\subsection{On the origin of UDG~32}\label{sec:UDG32_origin}

Based on the nature of the stellar filaments in NGC~3314A, two possible scenarios are viable for the origin of
UDG~32, assuming that these systems are physically connected.
If the stellar filaments in NGC~3314A result from RPS, the UDG might have formed from the gas clumps.
Alternatively, the UDG might originate from the stripped material (stars and gas) in the filaments, 
caused by a tidal interaction of the parent galaxy with other group members in the past.


The possible origin of UDGs from the star-forming clumps of gas in the jelly-fish galaxies has recently been  suggested by 
 \citet{Poggianti2019}. To date, no UDGs have been  {indisputably identified within the filaments of a jelly-fish galaxy}. 
 The stellar mass of  UDG~32 (M$\sim 10^{8}$~M$_{\odot}$) is consistent with the stellar mass 
 range ($ 10^{5} - 10^{8}$~M$_{\odot}$) of the clumps in jelly-fish galaxies \citep{Poggianti2019}.
{  If the UDG originated from RPS, we would expect a young stellar age 
\citep[$10^7$ up to $10^9$ yr,][]{Poggianti2019}, and bluer colours than those observed for  UDG~32.}
{  However, according to \citet{Poggianti2019},  star-forming clumps are affected by moderate extinction due to dust 
($A_V\sim0.5$~mag). 
In NGC~3314A, \citet[][]{Keel2001} found that the extinction in isolated and well-defined dust clumps reaches 
{  $A_B=0.4$~mag  ($A_V=0.3$~mag)}, being  even larger ($A_B>1$~mag) in the interspersed dusty arms. 
Therefore, if  UDG~32 formed by the materials in the filaments, dust
could be present and their intrinsic colours might be bluer. }
Gravitationally bound systems, with a baryonic mass similar to that of dwarf galaxies, could form in the tidally stripped material
during galaxy interactions: this is the mechanism proposed for the formation of TDGs, observed in several interacting systems 
\citep[][]{Duc2012,Ploeckinger2018}.
If the interaction of gas-rich galaxies is recent or ongoing, the TDGs are still star forming and should be blue.
In the sample of TDGs studied by \citet[][]{Duc2014}, two galaxies have $R_e$ and $\mu_0$ consistent with being UDGs 
(NGC5557-E1 and NGC5557-E2). Their stellar masses ($1.2\times 10^8$~M$_{\odot}$ and $0.15\times10^8$~M$_{\odot}$, respectively), 
are comparable with that of  UDG~32 in Hydra I. On the other hand, they have bluer colours 
(g-r$\sim$0.2-0.4 mag) with respect to  UDG~32, significant HI emission, and ongoing star formation.

{  Assuming that NGC~3314A might have weakly interacted in the past with a cluster member, which induced the formation 
of the stellar filaments, UDG~32 might originate from the material in the tidal tails of gas and stars.}
This formation mechanism for UDGs has been suggested by \citet[][]{Bennet2018}, {and recently revisited by \citet[][]{Jones2021}}. 
The newly discovered UDGs they studied are clearly associated with stellar streams connected to the parent galaxy, 
which could result from past galaxy interaction in the group. 
Both UDGs in that study are very diffuse, with $R_e$ (2.60 and 2.15 kpc), and faint 
($\mu_{0,g}=26-27$~mag/arcsec$^2$), have reddish colors ($g-r=0.4-0.5$~mag) and a low UV emission, 
suggesting the absence of active star formation. All the above 
quantities and properties are quite similar to those observed in  UDG~32. 

{  As pointed out by \citet[][]{Jones2021}, the presence of GCs in UDGs might be a key quantity to disentangle 
between formation scenarios. UDGs formed as TDGs and from the RPS gas, should not host typical, old GCs. 
The possible presence of a few {  old} GCs in UDG~32 (see Sec.~\ref{sec:color}), with $N_{GC}$ similar to that found
in other UDGs in the cluster \citep[][]{Iodice2020c}, would reconcile 
with the possibility that this system is in projection on  top  of  the  stellar  filaments.
It is worth noting that the number of GCs detected in UDG~32 is still consistent with zero, 
although this represents a lower limit.
However, in the tidal stripping scenario also proposed here for the origin of UDG~32, the GCs could be tidally 
stripped together with the stars. Therefore, it remains a viable formation path for this UDG.
The formation mechanism for UDG~32 is unlikely the same as that of UDGs hosting rich GC systems \citep[][]{Beasley2016,Forbes2020a}, unless the tidal stripping removed also substantial fraction of its GCs \citep[][]{Bellazzini2020}.}

\section{Conclusions and perspectives}

In this letter, we reported the discovery of a UDG, named UDG~32, that lies within the 
{extended} faint stellar filaments of the spiral galaxy NGC~3314A, at $\sim40$~kpc from the galaxy centre. 
UDG~32 is one of the faintest and most diffuse galaxies 
found in the cluster.
We investigated the hypothesis that UDG~32 might have formed from the stellar and/or gas 
material in the filaments.
 



To our best knowledge, the detection of the faint system of stellar filaments extending over $\sim50$~kpc 
towards SW from NGC~3314A is a new discovery from deep VST images of the Hydra~I cluster. 
Their origin has still to be explained. 
We addressed the possibility that the stars in the filaments could  
{have formed in the RPS gas} or, alternatively, the spiral disk of NGC~3314A could have been tidally 
distorted through an interaction with a cluster member.
The above scenarios could provide two possible formation channels for UDG~32: 
it might have built up from one of the gas clumps resulting from RPS, 
or from the tidal material of interacting galaxies. 
{  The rather red colour ($g-r=0.54\pm0.14$~mag) of  UDG~32, similar to that in the disk of NGC~3314A, an age older than 1~Gyr,
 and the possible presence of few old GCs would reconcile with a moderately evolved stellar content 
and, therefore, with the tidal formation scenario.} 
Differently from the few other UDGs known in the literature and associated 
with tidal features, the stellar filaments in NGC~3314A are the most extended LSB structure where a UDG is clearly detected.

The possibility that UDG~32 is a foreground or background UDG falling in that region of the cluster 
cannot be ruled out.
{  Follow-up spectroscopy is required to confirm the cluster membership of the newly discovered UDG~32, 
and thus its physical association with NGC~3314A, and to unveil the nature of the stellar filaments. 
To probe the physical association of the UDG~32 with the filaments in NGC3314A a homogenous map of the stellar
and gas kinematics is needed. The velocity of stripped gas in RPS galaxies can differ from the systemic velocity of the parent 
galaxy up to $\sim1000$ km/s \citep[see][]{Bellhouse2017}. Therefore, given the location of the UDG at the outer edge of the 
stellar filaments, its redshift should be consistent with the possible radial velocity gradient inside the filaments, from the 
disk of NGC~3314A outwards. From stellar and gas kinematic maps we would also be able to disentangle the RPS from the tidal 
origin of the stellar filaments in NGC~3314A. In fact, RPS affects only the gas, leaving the stellar components unperturbed \citep[see][]{Poggianti2017,Gullieuszik2017}.} 

To date, there are no detected UDGs formed in RPS filaments, therefore, if confirmed, this would be the first case.
The structure of NGC~3314A could be even more complex, where tidal forces might have act in the past, while 
RPS is still ongoing. This is the scenario proposed to account for the optical and HI structure in NGC~1427A, 
member of the Fornax cluster \citep[][]{Lee-Waddell2018}. 
The new WALLABY data for the Hydra I cluster \citep[][]{Wang2021}
will help to disentangle the nature of the stellar filaments in NGC~3314A.  






\begin{acknowledgements}
{  We thank the anonymous referee for his/her useful suggestions that helped to improve the paper.} 
EI wishes to thank Paolo Serra for extensive discussions. 
MC acknowledges support from MIUR, PRIN 2017 (grant 20179ZF5KS).
We acknowledge support from the VST INAF funds.
GD acknowledges support from CONICYT project Basal AFB-170002.
CS is supported by an `Hintze Fellow' at the Oxford Centre for Astrophysical Surveys, which is funded through generous support 
from the Hintze Family Charitable  Foundation.
\end{acknowledgements}

 \bibliographystyle{aa.bst}
  \bibliography{Hydra}

\begin{appendix}

\section{Archival data for NGC3314AB}\label{sec:arc_data}
\begin{figure*}
	\includegraphics[width=9cm]{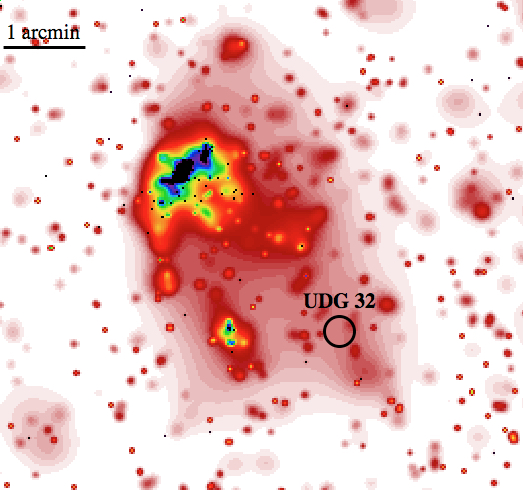}
	\includegraphics[width=9cm]{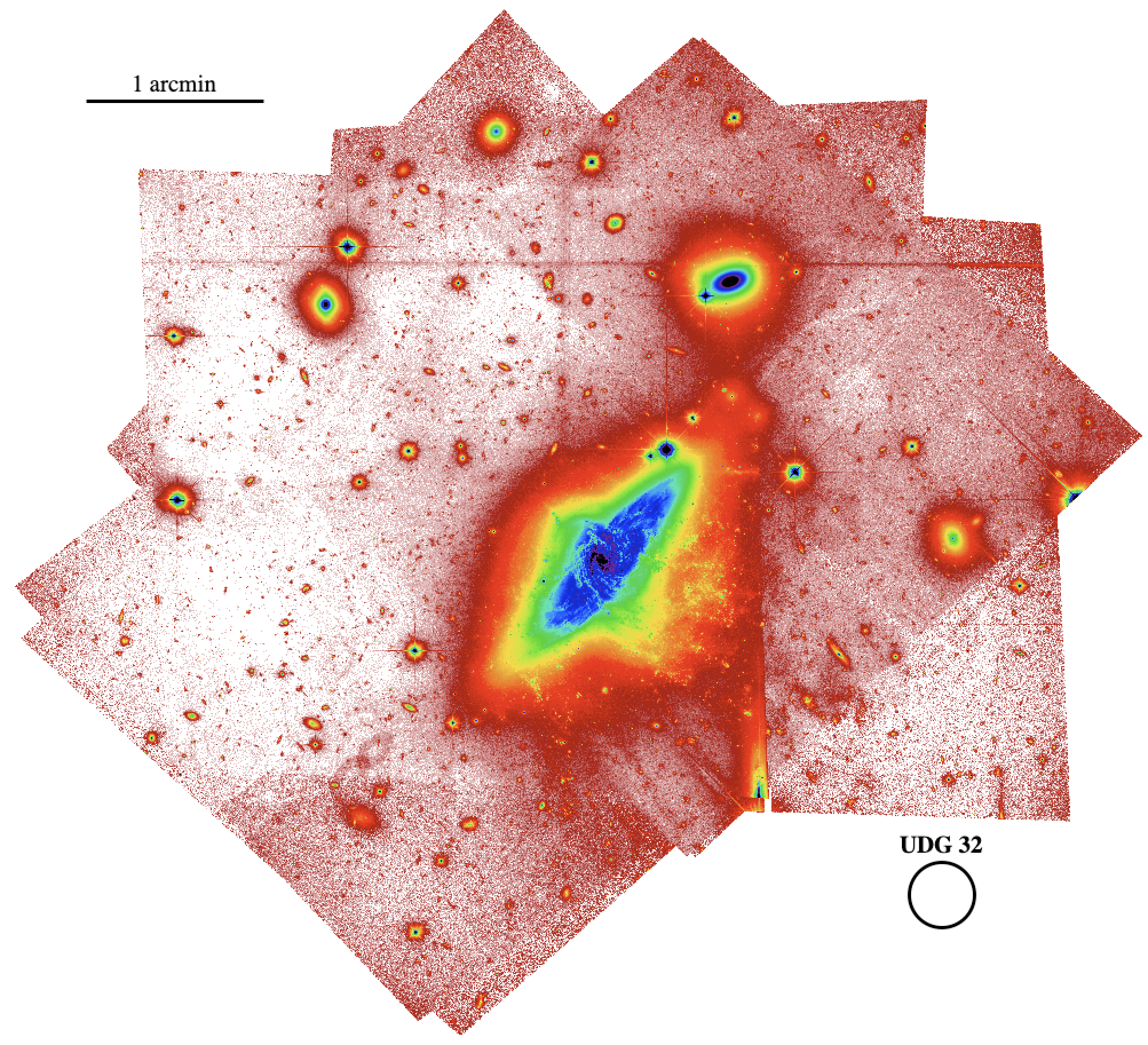}
	\caption{GALEX FUV emission in the region centred on NGC~3314AB of $6.44\arcmin \times 6.04\arcmin$ (left panel) and ACS/HST stacked mosaic ($\sim 5.28\arcmin\times3.67\arcmin$) of NGC~3314AB in the F606W filter (right panel). 
	{  The FUV image has been treated with {\tt ASMOOTH} \citep[][]{Ebeling2006}, 
selecting a S/N above the background of $\tau_{min}$=1.5.}
	The region of UDG~32 (marked with the empty black circle in both panels) is outside the ACS/HST mosaic.}
	\label{fig:UV_HST}
\end{figure*}

{  In Fig.~\ref{fig:UV_HST} are shown the ACS/HST and GALEX archival data for NGC~3314AB.
The ACS/HST mosaic is in the F606W filter. It covers a total area of $5.28\arcmin\times3.67\arcmin$ and has a total integration of 18hrs.
The GALEX data cover an area of $\sim1$~deg$^2$ and the FUV emission has a total integration time of 261 sec.}



\section{Surface photometry and structural parameters for UDG~32}\label{sec:detection}

{  In this section we describe the main steps adopted to derive the surface photometry and the structural parameters for UDG~32.
{  As described by \citet[][]{Iodice2020c}, the observing strategy and data reduction 
adopted for Hydra I are optimised to study galaxies and features
in the LSB regime. The long integration times (2.8 hours and 3.22 hours in the $g$ and $r$ bands, respectively) ensured reaching integrated magnitude sensitivities\footnote{Derived as the flux corresponding to $5 \sigma$, with the RMS of the background $\sigma$ estimated over an empty area of 1 arcsec.} of $m_g=28.6\pm0.2$~mag and $m_r=28.1\pm0.2$~mag.} 
The observations used in this work were 
acquired with the step-dither observing strategy, which guarantees an accurate estimate of 
the  sky  background \citep[see e.g.][]{Iodice2016,Venhola2017}. Therefore, the final reduced mosaics for 
the Hydra I cluster are already sky-subtracted.
In addition, since in the area of this cluster there is a bright (7th-magnitude) foreground star on the NE side 
of the cluster core, 
during the data acquisition we took special care to put this star always in one of the two wide OmegaCam  gaps,
thereby reducing the scattered light. 
The residual light from this bright star has been modelled and subtracted from the mosaic, in both bands.
The light distribution of the second brightest star in the field, located SE the core, is also modelled and subtracted from the parent
image \citep[see Fig.1 in][]{Iodice2020c}.
On the residual image, we have analysed the SE region of the cluster, where NGC~3314A and UDG~32 are located (see Fig.~\ref{fig:mosaic}).

The main steps of the photometry, to derive surface brightness profiles, colors profiles and integrated magnitudes and colors
are extensively described in many VEGAS papers \citep[see][and references therein]{Iodice2016, Spavone2017, Iodice2019}.
In short, we estimated the residual sky fluctuations around each galaxy of the cluster out to the most extended radius and, therefore,
the corresponding limiting radius ($R_{lim}$) where the galaxy light blends into this residual sky. 
During this process, all bright (background and foreground) sources, including the artefacts derived from the modelling of the brightest
stars in the field, are masked and excluded from the analysis.

In the case of UDG~32 we proceeded in two steps. 
Firstly, we estimated the average background at large distances from the center of
NGC~3314A, {  which starts to dominate at $R_{lim} \geq 120$ arcsec,} and the RMS scatter is $\sim 20\%$ and
$\sim 15\%$ in the g and r bands, respectively.
Since UDG~32 is located on top of the stellar filaments of NGC~3314A, the diffuse light from them also contributes 
to the integrated light of this object. Therefore, in order to account for this contribution, 
a local background value around UDG~32 is required.
{  This is computed in a circular annulus of $20 \leq R \leq 40$~arcsec centred on the UDG 
(see lower-right panel of Fig.~\ref{fig:mosaic}). 
The local background is then adopted to estimate the integrated magnitudes and colors, and 
the surface brightness profiles derived for UDG~32, reported in this work.}

The background value derived around NGC~3314A is used to compute the integrated magnitudes of the regions in the stellar
filaments (see Sec.~\ref{sec:3314}). In fact, the local background around each of them cannot be considered, since it 
coincides with the flux that we aim at measuring.}

\begin{figure*}
	\includegraphics[width=9cm]{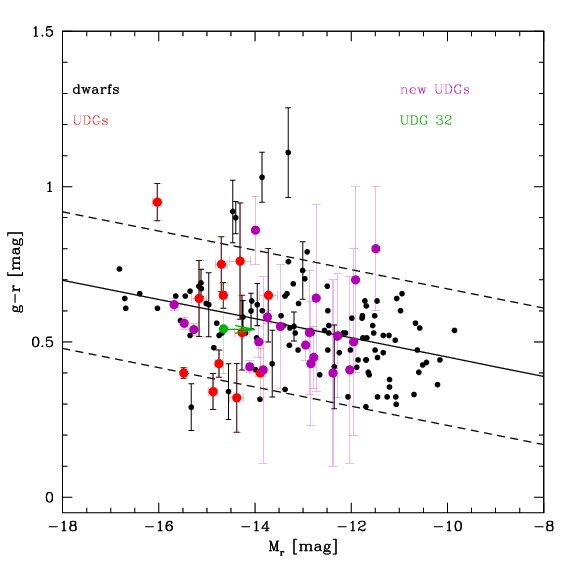}
	\includegraphics[width=9cm]{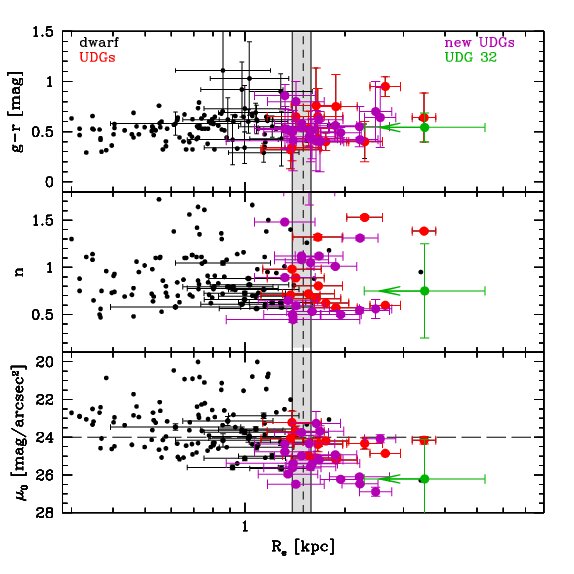}
	\caption{Left panel: Color-magnitude relation for the full sample of dwarf galaxies (black points) 
	detected in the VST Hydra I mosaic \citep{Iodice2020c} and from \citet{Misgeld2008}.
	Red filled circles indicate the UDG candidates found by \citet{Iodice2020c}. 
	The solid black line is the CM relation for the Hydra I cluster early-type galaxies derived by \citet[][dashed lines indicate the 1$\sigma$ scatter]{Misgeld2008}. { The magenta filled circles are the new UDGs detected in the cluster.} 
	The green point indicates UDG~32.
	Right panels: Structural and photometric parameters for the newly discovered UDGs (filled magenta circles) and for  
	UDG~32 in the filament of NGC~3314A (filled green circles) as a function of the effective radius. 
	The UDGs previously presented by \citet{Iodice2020c} are marked with red circles. The UDG definition criteria, $R_e\geq1.5$~kpc and $\mu_0\geq24$~mag/arcsec$^2$ \citep{vanDokkum2015}, are shown by the dashed lines. The black points are dwarf galaxies in Hydra I from \citet[][]{Misgeld2008} and \citet{Iodice2020c}. The vertical shaded region indicates the range of uncertainty on $R_e$. }
	\label{fig:corrRe}
\end{figure*}

\begin{figure*}
   \includegraphics[width=18cm]{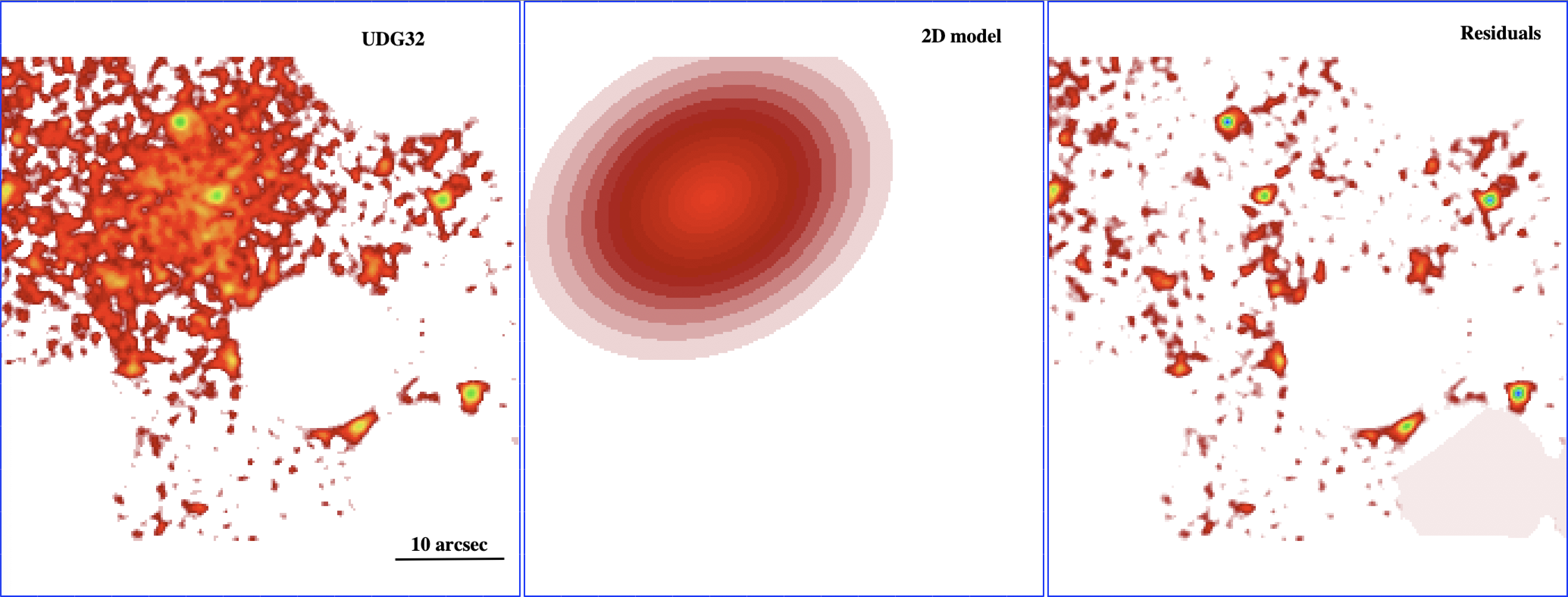}
	\caption{ Results from the 2D fit of the light distribution for the UDG~32, in the $g$ band, 
	with the {\sc GALFIT} tool \citep{Peng2010}. The parent image is shown in the the left panel. The 2D model and residuals are
	shown in the middle and right panels, respectively. Bright stars, foreground and background objects are masked and excluded from 
	the fit.}
	\label{fig:galfit}
\end{figure*}

{ Using the same analysis and technique presented by \citet[][]{Iodice2020c}, for UDG~32, 
and for all the new LSB candidates detected in the Hydra~I cluster as well, we have  
\textit{(i)} measured the total magnitudes, {  as aperture photometry for $R\leq R_{lim}$}, 
and the average $g-r$  colours, and 
\textit{(ii)} derived the structural parameters by fitting
the 2-dimensional light distribution, in the $g$ band, with {\sc GALFIT}  \citep{Peng2010}, 
adopting a single Sersic function. 
In Fig.~\ref{fig:corrRe} we show the color-magnitude relation and the structural parameters for the whole sample of UDGs, including the UDG~32, and dwarf galaxies in the Hydra~I cluster. 
For UDG~32, as for all UDGs in the sample, we have adopted the cluster distance of $51\pm6$~Mpc.
Assuming that the UDG~32 is physically associated with NGC~3314A, which has a distance of $\sim 37.3$~Mpc \citep[][]{Christlein2003},
we obtain $R_e=2.5\pm1$~kpc. Fig.~\ref{fig:corrRe} shows that this estimate of $R_e$ is still consistent 
with the average values
derived for the other UDGs in the sample, which have  {similar faint} surface brightness and large 
sizes $2\leq R_e\leq 3$~kpc. 
For UDG~32, the 2D model and residuals obtained by GALFIT are shown in Fig.~\ref{fig:galfit}. 
}

\section{2D number density distribution of UDGs in Hydra~I}\label{2D_density}

\begin{figure*}
   \includegraphics[width=9cm]{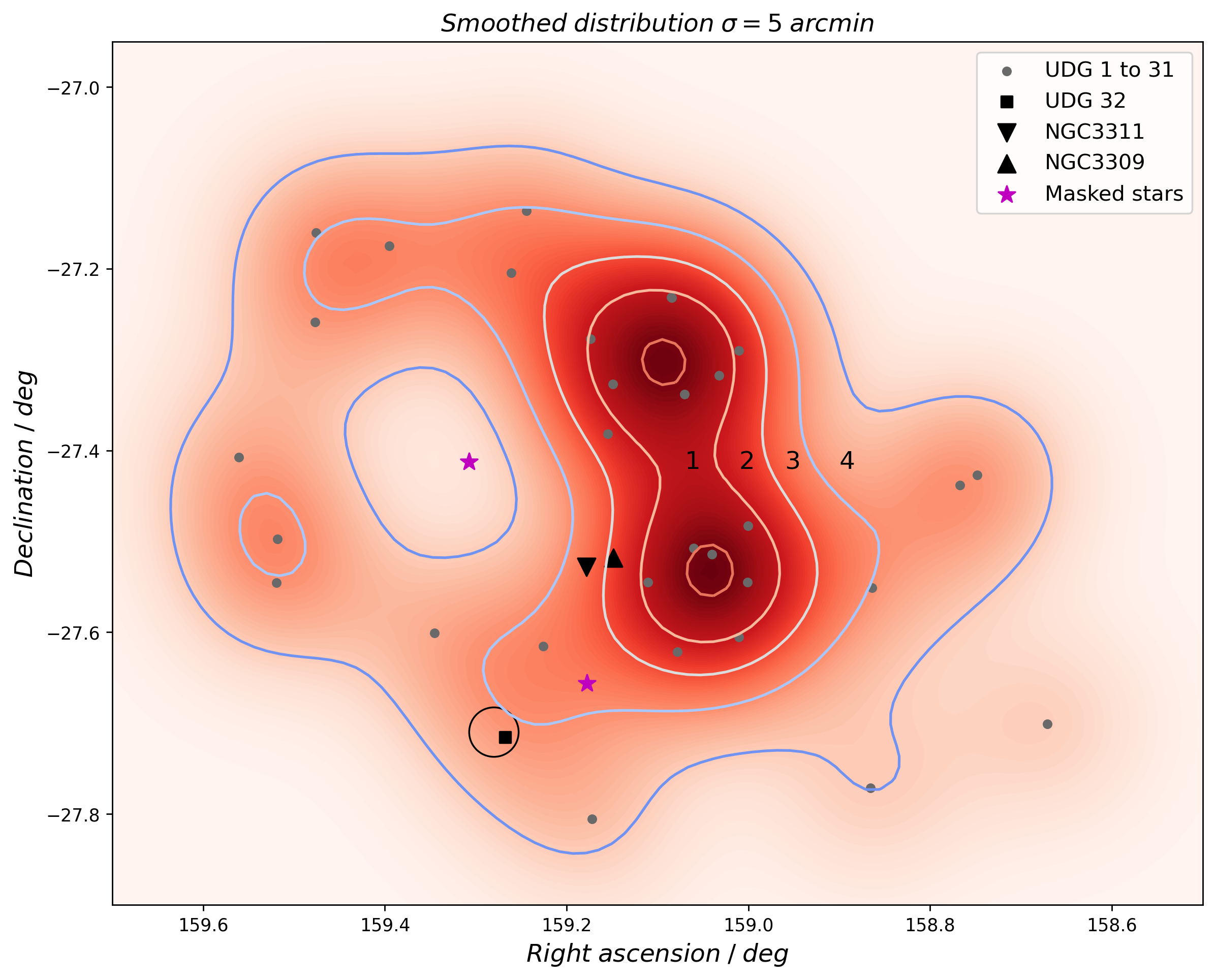}
   \includegraphics[width=9cm]{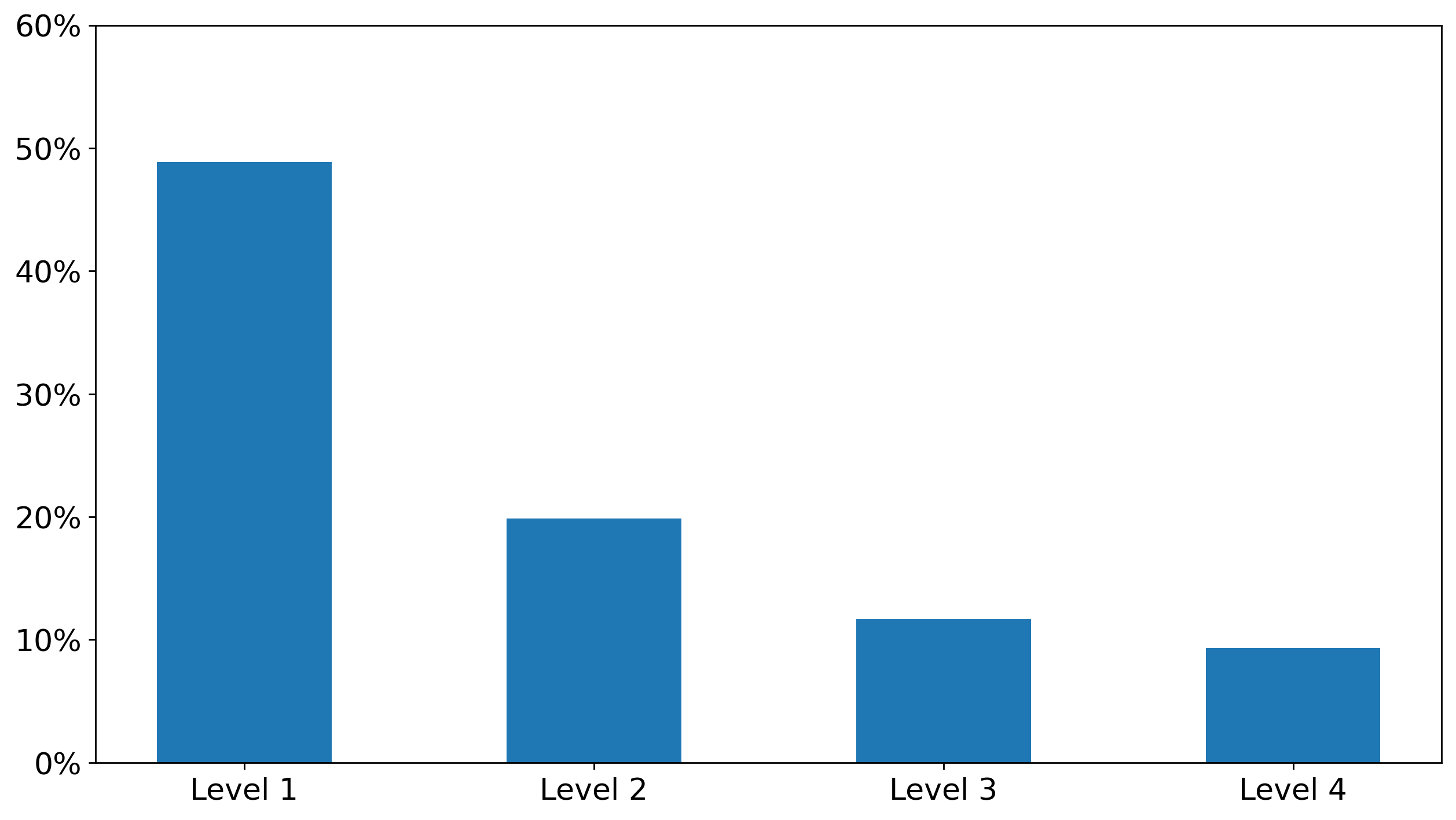}
	\caption{ Left panel: 2-dimensional distribution of UDGs in the Hydra~I cluster. 
	The peak of the UDG density distribution is close to the core of the cluster, which is where the two brightest cluster galaxies NGC~3311 and NGC~3309 are shown with the two black triangles.
	A lower number density is found at larger cluster-centric radii. 
	All UDGs are marked as grey circles. The black square and open black circle mark the position of the UDG~32 and the region of NGC~3314A. 
	The two brightest stars in the field are marked with magenta points. 
	The coloured solid lines represent the density contours. 
	Right panel: Probability histogram to find a UDG inside an area of 0.00232 deg$^2$ (assumed to cover the extension of NGC3314A and its stellar filaments, see text for details), as function of the UDGs number density.
	Each level of the UDGs number density is also shown in the left panel.}
	\label{fig:2Ddist}
\end{figure*}

{ Fig.~\ref{fig:2Ddist} shows the 2D number density distribution of all UDGs detected in Hydra I. 
The smoothed distribution is obtained convolving the galaxy distribution with a Gaussian kernel with a standard deviation of $\sigma = 5\;arcmin$.
This suggests that UDGs in Hydra I are not uniformly distributed: the UDGs distribution peaks close to the cluster core, and a lower number density is found at larger cluster-centric radii.
Based on this evidence, we derived the probability to find a UDG in a circle of fixed area as a function of the number density levels.
Therefore, we have defined a circle that covers the whole extension of NGC~3314A stellar filaments, with a radius of 0.0272 deg, and a total area of $A_{tail}= 0.00232$~deg$^2$, centred on R.A. = 159.28 DEC = -27.71 deg.
The probability to find a UDG within $A_{tail}$ is derived inside each density level as 
$N_{UDGs} \times Area_{tails}/Area_{total}$, where $A_{TOT}$ is the total area between the two contiguous contours.
The probability values are plotted in the right panel of Fig.~\ref{fig:2Ddist}.
In the region of the cluster where the UDG~32 is located, $A_{TOT}=0.275$~deg$^2$, and the total number of UDGs inside
this level is 11. Therefore, assuming that they are randomly distributed inside it, the probability in this region is $\sim9.3\%$.}

\section{Search for globular clusters around UDG~32}\label{sec:GC}

To detect candidate GCs around UDG~32 we have adopted the same procedure and assumptions to identify {  old} GCs as for the 
first sample of UDGs in Hydra~I, described by \citet[][]{Iodice2020c} and more extensively presented 
in \citet{cantiello2018,Cantiello2020}. 
In short, we run SExtractor on a large image cutout of 
$\sim10.5\arcmin \times 9.5\arcmin$, in the $g$ and $r$-band, where the model distribution of the UDG has been subtracted.
%
%
%
For each source, and in each band, we derived the automated aperture magnitude
($MAG\_AUTO$) to estimate the total magnitude of the source, and the aperture magnitude within 4 and 6 pixels diameter 
($MAG\_APER$) to estimate its $g-r$[6 pixels] colour 
and the concentration index (as $CI_X=MAG\_APER_X[4pixel]{-}MAG\_APER_X[6pixel]$), 
which is an indicator of source compactness \citep{peng11}.
%
%
To identify GC candidates, we select sources with:
{\it i)} $g$-band magnitude $23.5\leq m_g\leq 26$ mag, the expected range between the turn-over magnitudes (TOM) 
of the GC luminosity function (GCLF) and 3$\sigma_{GCLF}$ mag brighter \citep{villegas10};
{\it ii)} $0.25\leq g{-}r \leq 1.25$~mag; 
{\it iii)} SExtractor {  $CLASS\_STAR\geq0.4$};
{\it iv)} elongation (i.e. major-to-minor axis ratio) $\leq2$ in both bands; 
{\it v)} $CI_X$ within $\pm0.1$ mag of the sequence of local point sources. 


We find that the number of GC candidates over the $\sim10.5\arcmin \times 9.5\arcmin$ area around the 
UDG is $\sim 3.0 \pm 3$ arcmin$^{-2}$.
The total number of GCs ($N_{GC}$) has been derived within 1.5 $R_e$, adopting the approach suggested by  \citet[][]{vanDokkum2016}, and also within 3 and $5 R_e$, 
the latter being close to the upper limit for bound systems \citep{Forbes2017,Caso2019}. 
All estimates are corrected for the contamination of foreground stars, background galaxies, and 
possible intra-cluster GCs, {  which is $\sim 3$~sources/arcmin$^2$},
derived in the regions between $5\arcmin\leq R \leq 10\arcmin$ around the UDG. 
Since the photometry reaches roughly the TOM peak and assuming the GCLF 
is a Gaussian also in UDGs  \citep[e.g.][]{Rejkuba2012,vanDokkum2016}, 
we derived $N_{GC}$ as twice the background corrected GC density over the 3 and $5 R_e$ area of the UDG, times the 
area. The only difference with the $N_{GC}$ within 1.5 $R_e$ is that it is assumed 
that only half of the GC population is within 1.5 $R_e$, hence the total population 
is estimated as four times the background corrected GC density over 1.5 $R_e$ area, 
times the area.
Therefore, we find $0\leq N_{GC}\leq 6$ within all 
assumed radii. A similar value for $N_{GC}$ is found also using  {more stringent criteria}
($0.4\leq g{-}r \leq 1.1$~mag, CLASS\_STAR$\geq0.6$, $CI_X$ within $\pm0.05$ mag). 
In particular, we identify a compact source near the centre of the 
UDG (at $\sim3.5\arcsec$), with  colour, magnitude, and compactness properties close to what is expected
for GCs ($g-r\sim0.6$ mag, $m_g\sim24.9$, CLASS\_STAR$\sim0.9$ in both bands, and $CI_X$ 
consistent with the stellar sequence within $\leq0.05$ mag). 



\end{appendix}

\end{document}